# Estereoscópio de Wheatstone "Revival"

José J. Lunazzi, Milena C. França, Andrey da S. Mori
Instituto de Física - UNICAMP
lunazzi@ifi.unicamp.br

**Resumo**

Descrevemos a montagem de um estereoscópio com dois espelhos, idêntico ao primeiro da história, com a grande vantagem de usarmos imagens digitais em telas de cristal líquido. Com ele temos surpreendido ao público, que não imagina que pode ver 3D sem precisar de óculos específicos, nem que algo tão simples e antigo não seja bem conhecido. Agora que é comun poder colocar dois monitores num computador, a montagem se mostra simples.

**Introdução**

Vivemos uma época onde a imagem digital está cada vez mais em nosso cotidiano. A maioria das pessoas pode tirar uma foto com seu telefone celular e vê-la imediatamente. Os jovens de hoje não chegam a suspeitar como era o processo no século passado e retrasado: cuidados para não receber luz indesejada no filme fotosensível, revelação no escuro com líquidos que devíam estar bem conservados, controle de temperatura, lavagem, secagem, etc., para ter o resultado em um papel, o qual somente podía chegar a outro lugar se transportado. Duplicação, conservação, mudanças de formato, efeitos diversos, só se conseguíam com demorado trabalho de laboratório. Os princípios ópticos e os conceitos, no entanto, não mudaram: tamanho angular do campo (objetivas grande angular, normal, teleobjetiva e zoom nas câmeras profissionais), profundidade de focalização em função do tempo de exposição e sensibilidade, tempos mínimos para a foto não resultar borrada por movimento, efeitos de contraste entre regiões com muita e pouca luz, etc. As imagens atuais são já todas digitais e vêm progredindo na facilidade de obtenção (câmeras eletrônicas cada vez mais eficazes e baratas), qualidade de resolução (número de pixels), fidelidade na resposta à cor (propriedade que é menos conhecida do público) e na resposta às variações de intensidade da luz. Há evoluções também na nitidez da imagem, a qual está sendo quadruplicada com os sistemas chamados 4K. Temos a portabilidade (incorporação aos telefones de bolso, depósito em servidores gratuitos), possibilidades de edição (com cada vez mais e melhores programas sendo disponibilizados), e a entrada na era do vídeo. Mas há um avanço que ainda não chega a ser definitivo: o da imagem tridimensional, a chamada historicamente de 3D, e que hoje vai sendo melhor identificada como S3D (Stereo 3D), pois o termo original foi usurpado pela Computação Gráfica.

Sabemos que o mundo tem três dimensões espaciais, no entanto não conseguimos captá-las em nossas fotografias. As tentativas para o conseguir são tão antigas quase quanto a fotografia mesmo, senão mais antigas, pois os desenhistas estudam há séculos a perspectiva nas imagens. A diferença está em que um desenho ou uma fotografia não registram tudo da maneira que é visto em uma cena. Somente um holograma consegue isso, permitindo que os olhos focalizem alternadamente nos diferentes planos e converjam sobre os objetos escolhidos. Nas adaptações, mesmo as mais recentes e as que não precisam de óculos, obriga-se ao cérebro a ter de aprender a tolerar diferenças, algo que uma boa parte das pessoas, sobretudo as adultas, não consegue. Mesmo assim, o cinema 3D já é algo estabelecido na indústria, enquanto o uso doméstico ainda é incipiente. Pessoalmente, atribuimos a falta de avanço na utilização da tecnologia de visualização 3D à falta de atenção dada às possibilidades de criar as próprias imagens, pois a indústria fica alimentando somente o consumo das que ela própria produz.

No século passado a visualização em 3D foi popular bem no começo e alguns fotógrafos trabalharam na produção de imagens. No Brasil, temos registros do ano 1865[1]. Podia-se comprar câmeras e complementos para a fotografia 3D, chamada de estereoscópica e que se resume a realizar um par estéreo: imagens dos pontos de vista do olho esquerdo e o direito[2,3].

Mas nos anos sessenta, o cinema 3D e a fotografia 3D vão sendo menos produzidos pela indústria e somente grupos menores o continúam. No Brasil, foi utilizado em fotografia de casamentos[4] mas tecnicamente somente reaparece em 1989 como consequência da invenção da tela holográfica para luz branca[5,6], que teve as aplicações iniciais em vídeo por meio da técnica de par estéreo, realizando assim o primeiro filme tridimensional do país[7], que teve continuidade somente anos depois[8]. E também os primeiros desenhos de computador para visualização 3D (S3D). Não é difícil utilizar um desenho de computador para uma visualização S3D[9]. Se consegue por meio da escolha de dois pontos de vista que geram o par estéreo, como é amplamente usado hoje no cinema 3D. Neste campo, temos um brasileiro como seu melhor realizador[10].

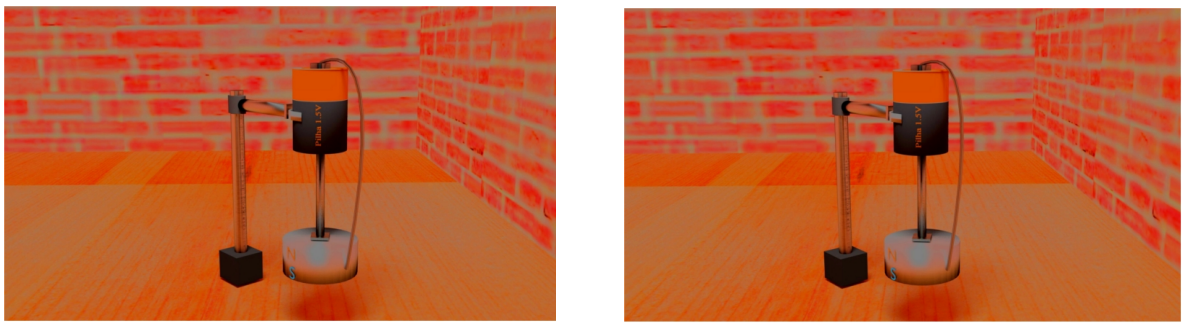

*Figura 1: Par estéreo desenhado por computador, gerado pela escolha de dois*

*pontos de vista horizontais deslocados. Embora pareçam iguais, a figura da esquerda mostra os elementos posicionados a respeito dos tijolos do fundo diferentemente da figura da direita. Realizado por Renán W.G. Miranda com o programa livre Blender 3D.*

Poderíamos, nesta época onde a tecnologia é toda desenvolvida por poucas empresas estrangeiras, fazer algo interessante relativo a imagem S3D? Pois sim, assim como uma câmera 3D pode ser desenvolvida(11) também podemos fazer um sistema de visualização. Neste trabalho reproducimos o primeiro visualizador para imagens 3D já criado e o aplicamos às imagens digitais.

**O estereoscópio de Wheatstone, condições para seu desenho**
O primeiro estereoscópio conhecido foi apresentado por Charles Wheatstone em 1838(12,13,14), ou seja, poucos anos depois da invenção da fotografia. O funcionamento é evidente e simples: dois espelhos em ângulo de noventa graus colocam-se enfrentados a cada olho do observador, o qual vê as fotografias (ou eventualmente desenhos) de um par estéreo que são localizadas paralelas e a cada lado, a noventa graus da visão dele, à esquerda e à direita (Fig.1).

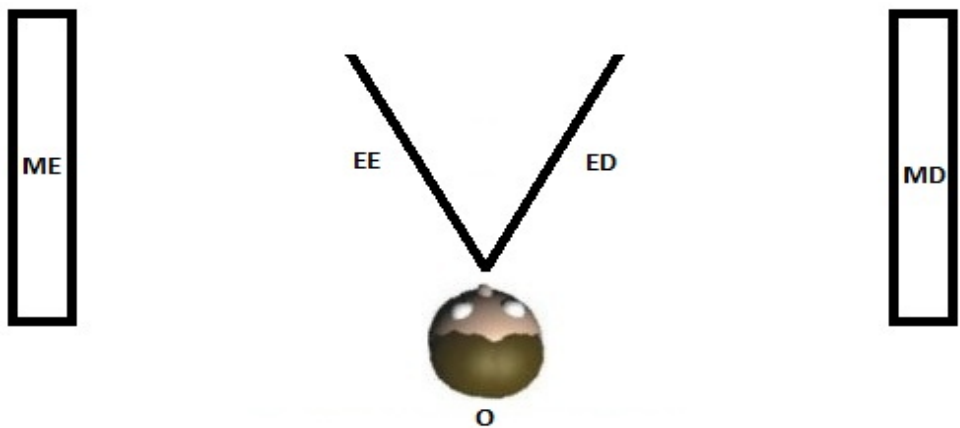

*Figura 2: Observador enfrentado a dois espelhos a $45^0$ dele, para ter em cada olho a visão de cada monitor correspondente. O: observador. EE, ED, espelhos para o olho esquerdo e direito. ME, MD, monitores colocados à esquerda e à direita.*

Ele foi necessário porque ter as fotografias simplesmente a frente não permite a separação visual para cada olho, algo que somente pode ser conseguido com algum treinamento para figuras de não mais de uns 6cm x 6cm. Nesse caso a pessoa pode colocar seus olhos em visão virtual para o infinito, e isso faz cada olho ver separadamente as figuras a curta distância, no par estéreo com colocação a esquerda da vista esquerda, e a direita da vista direita. Para algumas pessoas é mais fácil conseguir a sobreposição de imagens cruzando os olhos, nesse caso coloca-se o par estéreo invertido, com a vista esquerda à direita da vista direita. Para facilitar a tarefa, David Brewster introduziu em 1849 uma lente para cada olho poder focalizar de perto as imagens, resultando um modelo mais compacto e popular, muito utilizado por mais de um século. Esse modelo pode ser adaptado a nossa era digital, porém não é popular(15). Notemos que o de Wheatstone inverte especularmente cada imagem (letras, e consequentemente textos, ficam ininteligíveis; tal problema pode ser corrigido fazendo a inversão das imagens digitalmente, usando um editor de imagens como o GIMP) enquanto o de Brewster necessita ajustar a separação das lentes conforme a separação entre olhos do observador e cria problemas para quem usa óculos. Embora existam monitores de computador com capacidade para separar imagens por polarização(16), nos ocorreu utilizar o de Wheatstone por causa da dificuldade em fornecer óculos separadores para um grupo grande de pessoas, em eventos de férias, por exemplo. O custo desses elementos, que são importados, e manutenção, que envolve desinfecção, são fatores limitantes que conseguimos evitar dessa maneira. A falta de popularização e o alto preço colocado no país impedem o que seria ideal: que as pessoas fossem a eventos de imagem 3D levando seu próprio óculos.

**Resultados experimentais**

Fizemos o cálculo do tamanho mínimo necessário do espelho para poder ter os monitores de área útil de 21,5" de diagonal, 27 cm x 48 cm, à distância que nos pareceu apropriada, de 50 cm entre eles. Ele inclui os valores das distâncias mostradas no destaque da Fig.3:

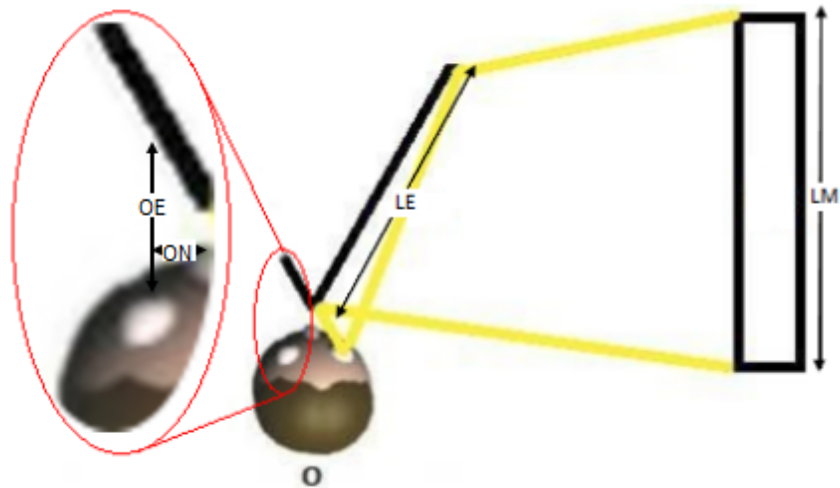

*Figura 3: Cálculo do tamanho mínimo de espelho. O observador (O) tem um olho à distância OE do espelho de largura LE e vê ao monitor de largura LM colocado à distância OM. ON seria a distância de seu olho em relação ao seu nariz.*

OM = 35 cm, distância mínima recomendada para focalização. OE é a distância do olho até o espelho. ON é a distância transversal entre o olho e o nariz. Fizemos o cálculo mas achamos que é uma complexidade desnecessária, sendo apenas um exercício de óptica geométrica não trivial. O deixamos de colocar aqui pois há um método prático: Para determinar o tamanho mínimo do espelho, estabeleça a distância OM, ou seja, fique com o rosto a uma distância OM do monitor. Pegue um pedaço de cartolina e coloque à frente do nariz, de maneira que forme um ângulo de 45 graus entre o rosto e a mesma. Mude o tamanho da cartolina de maneira que a cartolina tampe a visão de um olho. Se bloquear completamente a visão do monitor, tal tamanho de cartolina é suficiente para o espelho. Em nosso caso o espelho precisava ter um tamanho mínimo de 8 cm x 10 cm mas acabamos usando um maior (17 cm x 17 cm) por praticidade de realizar uma montagem sem suporte e para dar mais campo de posições verticais ao observador. Colocamos o conjunto sobre uma madeira de 95 cm de comprimento e sustentada pela base em um tripé firme para facilitar a posição do observador, que fica sentado (Fig. 4). Note-se que seria preciso precompensar a inversão direita-esquerda que os espelhos criam, o que poderia ser feito por meio de qualquer editor digital de imagens ou, mais prático talvez para o caso de vídeos, colocando espelho na frente das câmeras. Senão acaba sendo preciso trocar o par estéreo de lado, o esquerdo na direita e reciprocamente, o direito na esquerda, com o que a imagem fica invertida porém com relevo normal, mas sem poder fazer leitura de letras que viesse a ter nela. Poderia ter um jogo de espelhos do outro lado, para mais um observador? Este caso deixamos como um exercício interessante para o leitor.

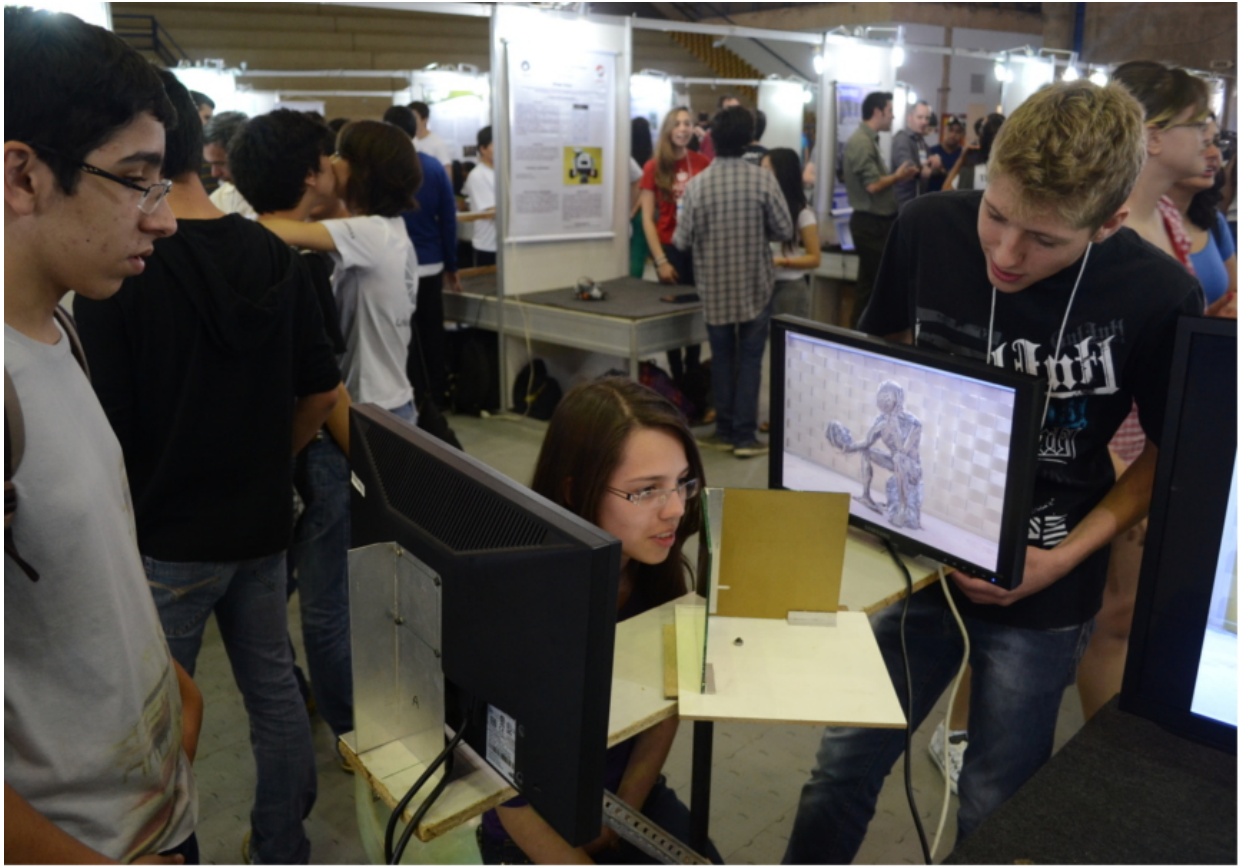

*Fig.4: Exibindo o estereoscópio em Feira de Trabalhos Técnicos.*

O estereoscópio foi exibido com sucesso para uma grande quantidade de público em três eventos(17,18,19), e em paralelo exibimos em uma TV 3D polarizada fotografias 3D realizadas por nós. O público era geralmente jovem e não identificamos que tivessem problemas na visualização 3D: Podemos pensar que quem assiste cinema 3D quando jovem tem mais facilidade para se adaptar à visualização estéreo?.

**Primeira transmissão 3D da internet?**
Por meio de duas câmeras web alinhadas para casar com a visão binocular (Figura 5 ) e usando em paralelo dois canais de vídeo pela internet, realizamos uma conexão em 3D real, ou S3D, que foi, por sua vez, visualizada sem precisar de óculos 3D, graças a nosso estereoscópio de Wheatstone. Não temos conhecimento de uma transmissão 3D pela internet, e menos provável parece ainda ser feita sem precisar de óculos. Teria Wheatstone imaginado isto? Em um artigo próximo iremos descrever em detalhe esta instalação mas no modo video-conferência, com projeção multimídia em tela.

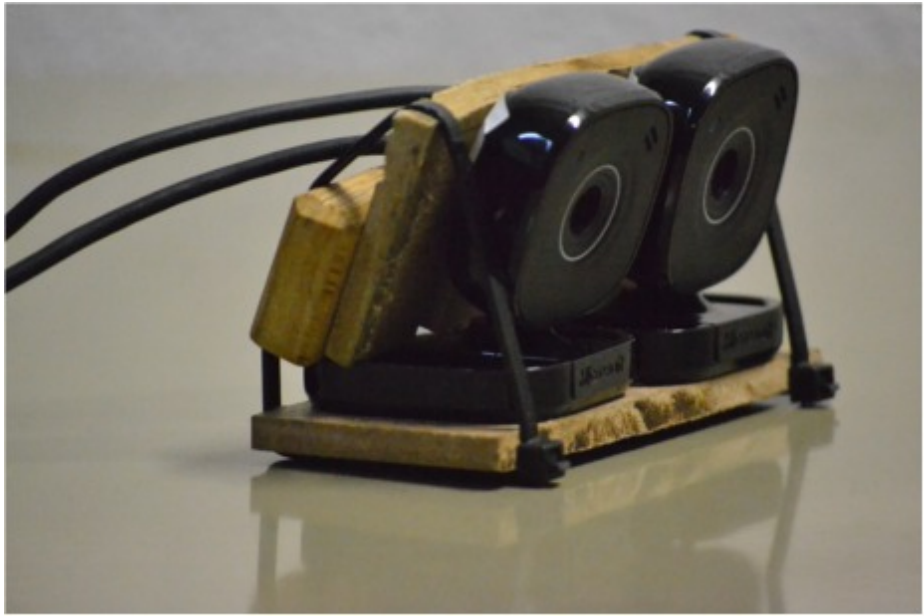

*Fig.5: Montagem de duas câmeras de computador para visualização 3D.*

**REFERÊNCIAS**

óculos chamados "passivos" (polarizados circularmente). Todas as telas polarizadas tem origem em um único fabricante, e, para evitar a visão parasita da imagem que não corresponde ao olho em questão, obrigam a manter uma distância mínima de quase 2m, o que não é prático para o usuário que fica perto.

17. III Feira de Trabalhos Técnicos do Colégio Técnico da UNICAMP-COTUCA, Ginâsio da UNICAMP, setembro de 2013. http://www.mostradetrabalhos.cotuca.unicamp.br/?id=10
18. Feira de Trabalhos de Iniciação Científica e Ensino do Instituto de Física da UNICAMP, 13 de novembro de 2013.
19. EXCUTE, 38a. Exposição Cultural e Tecnológica, de Santo André, Escola Técnica ETEC Jorge Street, S. Caetano do Sul-SP, 6 e 7 de dezembro de 2013.